\documentclass[aip,jap,amsmath,amssymb,
reprint, footinbib,unsortedaddress]{revtex4-1}
\usepackage{CJK}
\usepackage{dcolumn}
\usepackage{graphicx}
\usepackage{amssymb}
\usepackage{bm}
\usepackage{dcolumn}
\usepackage{bm}

\begin{document}
\begin{CJK*}{SJIS}{MS Mincho} 

\title{Thermal Transport Properties of Graphene-Based F$\mid$S$\mid$F Junctions}

\author{Morteza Salehi}\thanks{\textbf{Current address}: \textit{Department of Physics, Sharif University of Technology, Tehran 11155-9161, Iran.}}

\author{Mohammad Alidoust}\thanks{\textbf{Current address}: \textit{Department of Physics, Norwegian University of Science and
Technology, N-7491 Trondheim, Norway.}}

\author{Yousef Rahnavard}\thanks{\textbf{Current address}: \textit{Institute for Theoretical Physics, Technical University Braunschweig, D-38106 Braunschweig, Germany.}}

\author{Gholamreza Rashedi}

\affiliation{Department of Physics, Faculty of Sciences, University of
Isfahan, Hezar Jerib Avenue, Isfahan 81746-73441, Iran}

\date{\today}
\begin{abstract}
We present an investigation of heat transport in gapless
graphene-based Ferromagnetic /singlet Superconductor/Ferromagnetic
(FG$\mid$SG$\mid$FG) junctions. We find that unlike uniform increase
of thermal conductance vs temperature, the thermal conductance
exhibits intensive oscillatory behavior vs width of the sandwiched
$s$-wave superconducting region between the two ferromagnetic
layers. This oscillatory form is occurred by interference of the
massless Dirac fermions in graphene. Also we find that the thermal
conductance vs exchange field $h$ displays a minimal value at
$h/E_F\simeq 1$ within the low temperature regime where this finding
demonstrates that propagating modes of the Dirac fermions in this
value reach at their minimum numbers and verifies the previous
results for electronic conductance. We find that for thin widths of
superconducting region, the thermal conductance vs temperature shows
linear increment \textit{i.e.} $\Gamma\varpropto T$. At last we
propose an experimental set-up to detect our predicted effects.

\end{abstract}
\maketitle

\end{CJK*}

\section{\label{sec:intro}Introduction}
In the recent years monoatomic graphite layers and structures
containing this layers has attracted so much attentions
theoretically and experimentally to itself
\cite{beenakker1,beenakker2,beenakker3,beenakker4,beenakker5,beenakker6,beenakker7,Geim1,Geim2,Geim3,Geim4,Geim5}.
The monoatomic graphite layer is called graphene and naturally
treated as a two dimensional system.
 This two dimensional artificial system made
by Novoselov \textit{et al} \cite{Novoselov,Novoselov2}. In
graphene, low-excitation electrons follow the Dirac equation for
their behaviors in various conditions and consequently Dirac
equation can predict properties of the system \cite{Wall}. Graphene
exhibits very interesting properties that from one side confirm some
of the predicted phenomena in relativistic quantum mechanics
\textit{e.g.} Klein's paradox and from other side its high mobility
and controllable Fermi energy in experiment make it very interesting
and suitable in laboratory and industry
\cite{berger,beenakker1,Bunch}.
 From application point of view, it is very important to
know the transport properties (charge, spin and thermal transport
properties) of the devices including graphene junctions
\cite{Bhattacharjee,Linder2,Linder,Saito,Trushin,Peres}.

Thermal and charge transport properties are so much related to each
other. Charge conductivity of the normal-superconductor (N/S)
junctions at first was discussed by Blonder, Tinkham, and Klapwijk
(BTK) \cite{BTK} theoretically and their results could show good
consistence with experiments. B-T-K take into account the
contribution of the Andreev reflection \cite{Andreev} in the
electronic transport of N/S junctions and use Boguliobov-de Genne
(BdG) formalism to obtain the charge conductance at low
temperatures. The BTK theory is limited to the clean regime of the
heterojunctions, while cases with high impurities are not within the
regime of validity. Bardas \textit{et al.} and Devyatov \textit{et
al.} \cite{bardas,devyatov} generalized the BTK model for the charge
transport and calculated the thermal current through the N/S
junctions. For metals at low temperatures the thermal conductivity
$\Gamma$  has a linear behavior with respect to temperature
\textit{i.e.} $\Gamma\varpropto T$, and their electronic
conductivity reaches to constant value. So the Wiedemann-franz law
is satisfied for them \cite{wiedemann}. By depositing a
superconducting electrode on a graphene substrate, the
superconducting correlations can leak into the graphene, due to the
proximity effect \cite{beenakker3}. Also ferromagnetism can be
induced into the graphene by doping or using an external field
\cite{Oleg,Son,Asano}.
\begin{figure}
\includegraphics[width=9.5cm]{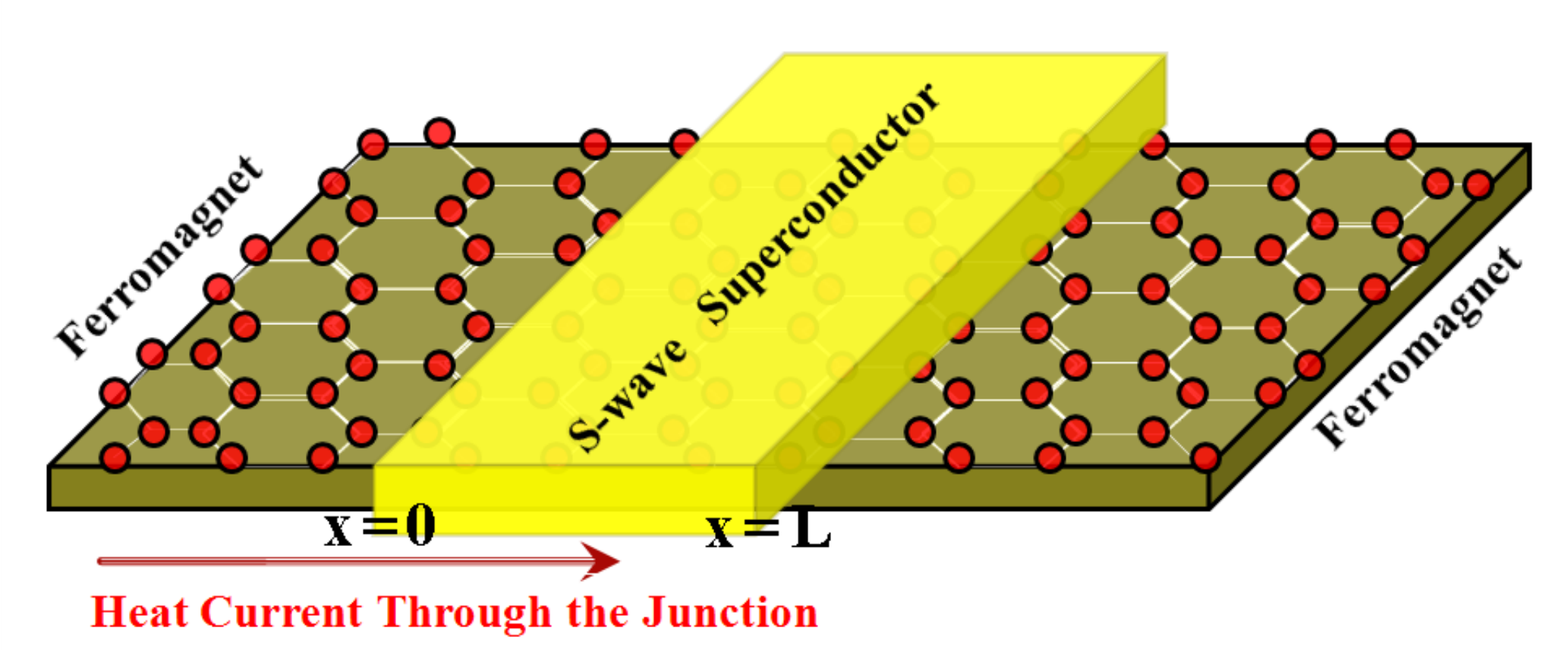}
\caption{\label{fig:model} (Color online) Schematic model for a
system of the sandwiched graphene-based $s$-wave superconducting
layer with width $L$ , between two uniform graphene-based
ferromagnetic substrates (FG/SG/FG). The ferromagnetic layers have
similar directions in both sides.}
\end{figure}

In this paper we utilize the Dirac Bogolubov de-Genne equation
and solve it for two dimensional systems including
FG/SG and FG/SG/FG graphene junctions in which
SG stands for the graphene-based $s$-wave superconductor
and FG stands for graphene-based ferromagnetic substrate.
 We use the generalized BTK formula for heat
transport through the junctions and investigate the thermal
transport properties of the FG/SG and FG/SG/FG junctions and as a
especial case, we assume the exchange field $h$, equal to zero for
approaching to normal case \textit{i.e.} NG/SG and NG/SG/NG. We find
that the thermal conductance of the junction exhibits an
\textit{intensive oscillatory shape}, damping simultaneously by
increasing the width of the superconducting layer. We find that the
thermal conductance shows an exponential
 increase vs. temperature for large widths of superconducting layer that
reflects the s-wave symmetry of the Dirac fermions inside the
graphene as was mentioned by BTK but \textit{we find that for thin
widths of superconducting layer within the tunneling regime, thermal
conductance $\Gamma$ is linearly proportional to temperature} T
\textit{i.e.} $\Gamma\varpropto$T.
 Also we find that the
thermal conductance vs. strength of the exchange field of the
ferromagnetic substrate has a minimum near $h\simeq E_F$ which this
value moves toward smaller values $h<E_F$ by increasing the
temperature.

\section{\label{sec:theory}Theory}
We consider a graphene-based ferromagnetic/superconductor junction
FG/SG which is placed in the $xy-$plane, and ideal interfaces
between ferromagnet and superconductor located at $x = 0$, are
perpendicular to the $x-$ axis. For investigating properties of the
mentioned system , one should solve the Dirac-Bogoliubov-de Gennes
equation by following the Refs. \cite{beenakker3,hakime}
\begin{eqnarray}
\left(
\begin{array}{ccc}
  H_0-\sigma h & \Delta \\
  \Delta^{*} & -(H_0-\bar{\sigma} h)
\end{array}
\right)\left( \begin{array}{ccc}
  u_\sigma \\
  v_{\bar{\sigma}}
\end{array}\right)=\epsilon_\sigma\left( \begin{array}{ccc}
  u_\sigma \\
  v_{\bar{\sigma}}
\end{array}\right), \label{BdG}
\end{eqnarray}
in which $H_{0}({\bf r})=-i\hbar
v_{F}(\sigma_x\partial_x+\sigma_y\partial_y)+U({\bf r})-E_{F}$,
where $\sigma_x$ and $\sigma_y$ are $2 \times 2$ Pauli matrices.
Also, $\Delta$ stands for order parameter of the superconducting
layer, and one can mention the order parameter of the system under
consideration as $\Delta({\bf r},T)=\Delta(T) \Theta(x)$, in which
$\Theta(x)$ is the well known step function, $\epsilon_\sigma$ is
the excitation energy of the Dirac fermions with respect to the
Fermi level, $h({\bf r})=h_{0}\Theta(-x)$ is the exchange field
energy of the ferromagnetic layer. Here $\sigma=\pm1$ stands for
spin-up and -down quasiparticles, and $\bar{\sigma}=-\sigma$.
$U({\bf r})$ shows the Fermi mismatch vector(FMV), and since through
out the paper we investigate heavily doping cases, we assume a large
value for mismatch potential in comparison with the Fermi energy
$E_F$ within the ferromagnetic region \textit{i.e.} $U(\bf r)=-U_0
\theta(x)$ and $U_0\gg E_F$. Both of the $u_\sigma$ and
$v_{\bar{\sigma}}$ include two components of sublattices in
hexagonal lattice of graphene.
 Thus each spinor in Eq.(\ref{BdG}) involves
four components, and for the electron-like excitations in the
ferromagnetic region ($x < 0$) we obtain:
\begin{equation}
\psi_{e,\sigma}^{\pm}(x,y)=\frac{1}{\sqrt{\cos{\alpha_\sigma}}}e^{(\pm
ik_{e,\sigma} x+iqy)}\left(\begin{array}{c}
     1 \\
     \pm e^{\pm i \alpha_\sigma}\\
     0\\
     0\\
   \end{array}\right),
\label{psiEF}
\end{equation}
and for hole-like quasiparticles:
\begin{equation}
\psi_{h,\bar{\sigma}}^{\pm}(x,y)=\frac{1}{\sqrt{\cos{\alpha_{\bar{\sigma}}}}}e^{(\pm
ik_{h,\bar{\sigma}} x+iqy)}\left(\begin{array}{c}
     0 \\
     0 \\
     1 \\
     \mp e^{(\pm i \alpha^{\prime}_{\bar{\sigma}})}\\
   \end{array}\right),\label{psiHF}
\end{equation}
 where $k_{e(h),\sigma}$ is component of the electron (hole)-like wave-vector, perpendicular to the interface and $q$ is parallel component of the wave vector which remains conservative
during the scattering process. The above appeared factors
$1/\sqrt{\cos(\alpha)}$ and $1/\sqrt{\cos(\alpha^{'})}$ guarantee
same particle current transport by the four wave-functions
\cite{beenakker3}.
 Also $\alpha_\sigma({\alpha^{\prime}}_{\bar{\sigma}})$ are injection angles of electron (hole)-like quasiparticles with respect to the axis normal to the interface ($x$-axis). They are defined as:
  \begin{eqnarray}
  \alpha_{\sigma}&=&\arcsin{\left(\frac{\hbar v_F q}{\epsilon+E_F+\sigma h}\right)},\\
    {\alpha^{\prime}}_{\bar{\sigma}}&=&\arcsin{\left(\frac{\hbar v_F q}{\epsilon-E_F+\sigma h}\right)},\\
      k_{e,\sigma}&=&\frac{\epsilon+E_F+\sigma h}{\hbar v_F}\cos\alpha_{\sigma} ,\\
    k_{h,\bar{\sigma}}
   &=&\frac{\epsilon-E_F-\sigma h}{\hbar
   v_F}\cos\alpha^{\prime}_{\bar{\sigma}}  .
   \end{eqnarray}
 In the superconductor region ($x>0$), wave function for the hole-like quasiparticles read as:
\begin{eqnarray}
\psi^{\pm}_{S,h}=e^{ i(\mp (k_0-i\chi) x+qy)}\left(\begin{array}{c}
     e^{-i\beta} \\
     \mp e^{-i(\beta \mp \gamma)} \\
      1\\
     \mp e^{(-i\gamma)}\\
   \end{array}\right), \label{psiSe}
\end{eqnarray}
and for the electron-like quasiparticles:
\begin{eqnarray}
\psi^{\pm}_{S,e}=e^{i(\pm (k_0+i\chi) x+qy)}\left(\begin{array}{c}
     e^{i\beta} \\
     \pm e^{i(\beta \pm \gamma)} \\
      1\\
     \pm e^{i\gamma}\\
   \end{array}\right), \label{psiSh}
\end{eqnarray}
where
\begin{equation}\label{beta}
\beta=\left(\begin{array}{c}
     \cos^{-1}(\frac{\epsilon}{\Delta_0}), \epsilon<\Delta_0 \\
     -i\cosh^{-1}(\frac{\epsilon}{\Delta_0}), \epsilon<\Delta_0 \\
   \end{array}\right),
   \end{equation}
\begin{eqnarray}
k_{0}&=&\sqrt{(\frac{U_0+E_F}{\hbar v_F})^2-q^2},\\
\chi&=&\frac{U_0+E_F}{k_0(\hbar v_F)^2}\sin{\beta},\\
\gamma&=&\arcsin{\frac{\hbar q v_F}{U_0+E_F}},
\end{eqnarray}
here $v_{F}$ is energy-independent Fermi velocity in graphene. We
define right going ($+x$-direction) and left going ($-x$-direction)
quasi-particle wave-functions with plus and minus signs
\textit{e.g.} $\psi^{+}$, $\psi^{-}$ respectively. In the mean field
approximation, we assume high doping regime $U_0+E_F \gg \Delta_0$
\cite{beenakker3}. It is clear that during the scattering process
from the interface, $q$ component of parallel wave vector and the
energy of quasiparticles are constant (elastic scattering). The wave
functions of the moving quasiparticles must satisfy the boundary
conditions at the interface between ferromagnet and superconductor
as in Ref. \cite{beenakker3}. For FG/SG junction the boundary
condition reads as:
\begin{equation}
\psi_{e,\sigma}^{+}+r_{N,\sigma} {\psi}_{e,\sigma}^{-}+r_{A,\sigma}
{\psi}_{h,\bar{\sigma}}^{-}=t_{e,\sigma}{\psi}_{S,e}^{+}+t_{h,\sigma}
{\psi}_{S,h}^{-}. \label{coeffeicent1}
\end{equation}
For the FG/SG/FG structure that schematically is shown in Fig.
\ref{fig:model}, however the boundary conditions at $x=0$ and $x=L$
respectively are:
\begin{equation}
\psi_{e,\sigma}^{+}+r_{N,\sigma} {\psi}_{e,\sigma}^{-}+r_{A,\sigma}
{\psi}_{h,\bar{\sigma}}^{-}=a_1
{\psi}_{S,e}^{+}+a_2{\psi}_{S,h}^{-}+a_3{\psi}_{S,e}^{-}+a_4{\psi}_{S,h}^{+},
\label{coeffeicentFSF1}
\end{equation}
and
\begin{equation}
t_{e,\sigma}{\psi}_{e,\sigma}^{+}+t_{h,\sigma}
{\psi}_{h,\bar{\sigma}}^{-}=a_1
{\psi}_{S,e}^{+}+a_2{\psi}_{S,h}^{-}+a_3{\psi}_{S,e}^{-}+a_4{\psi}_{S,h}^{+}.
\label{coeffeicentFSF2}
\end{equation}
\begin{figure}
\includegraphics[width=8.5cm]{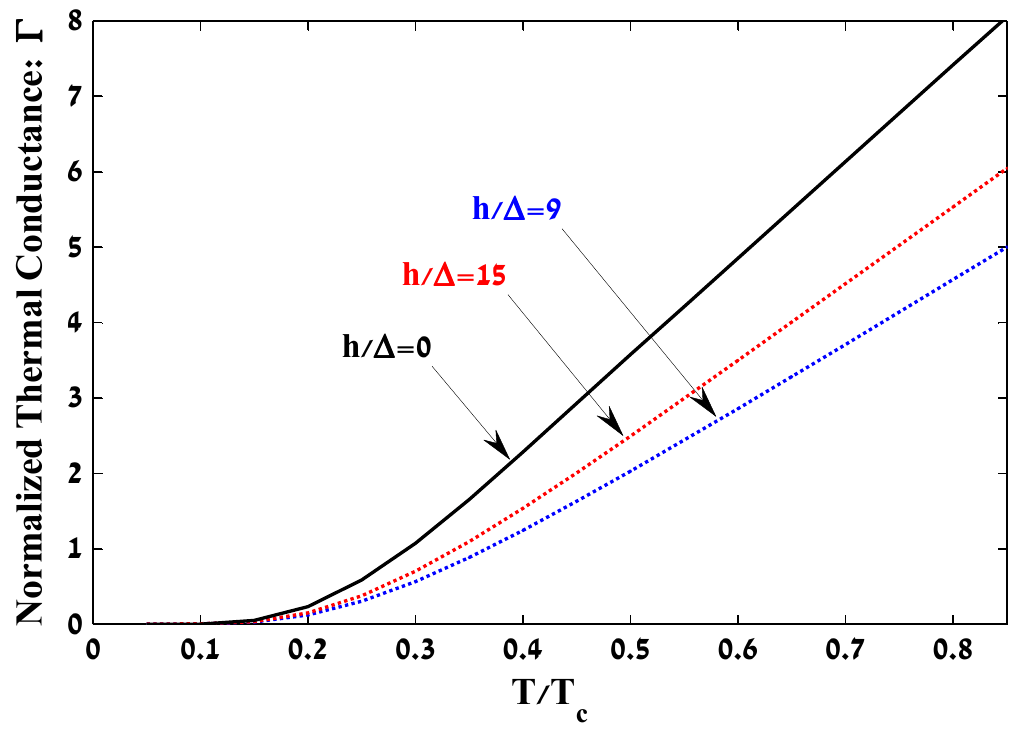}
\caption{\label{fig:KapavsTSF} (Color online) The normalized
thermal conductance of single FG/SG junction vs temperature for
three values of the ferromagnetic sheet's exchange field.}
\end{figure}
Here, $r_{A,\sigma}$ is amplitude of the Andreev reflection,
$r_{N,\sigma}$ is amplitude of the normal reflection, $t_{e,\sigma}$
and $t_{h,\sigma}$ are amplitudes of the electron-like and hole-like
quasiparticle's transmission, respectivley. Substituting the wave
functions into the above boundary conditions, we calculate the
coefficients in Eq. (\ref{coeffeicent1}), for FG/SG and FG/SG/FG
junctions. Then we now are able to obtain the probability of the
Andreev reflection ($R_{A,\sigma}=|r_{A,\sigma}|^2$) and normal
reflection ($R_{N,\sigma}=|r_{N,\sigma}|^2$). As seen in Fig.
\ref{fig:model} interfaces are normal to the $x$-axis and
superconductor region is between $x=0$ and $x=L$, so
the $x$-dependent order parameter can be written as $\Delta(x)=\Delta_0\theta(x)\theta(L-x)$.\\
The normalized thermal conductance $\Gamma=\Gamma^{'}/\Gamma_0$ is
given as follow\cite{linder3,bardas}:
\begin{eqnarray}
\nonumber\Gamma^{'}&=\Gamma_0&\sum_{\sigma=\uparrow\downarrow}\int_{0}^{\infty}\int_{-\pi/2}^{\pi/2}dE
d\alpha_{\sigma}\cos(\alpha_{\sigma}) \{1-\mid
r_{N,\sigma}(E,\alpha_{\sigma})\mid^{2}\\
&-&\mid r_{A,\sigma}(E,\alpha_{\sigma})\mid^{2}\}\frac{E^2}{T^2
\cosh^{2}(\frac{E}{2T})}, \label{Gamma}
\end{eqnarray}
\begin{figure}
\includegraphics[width=8.5cm]{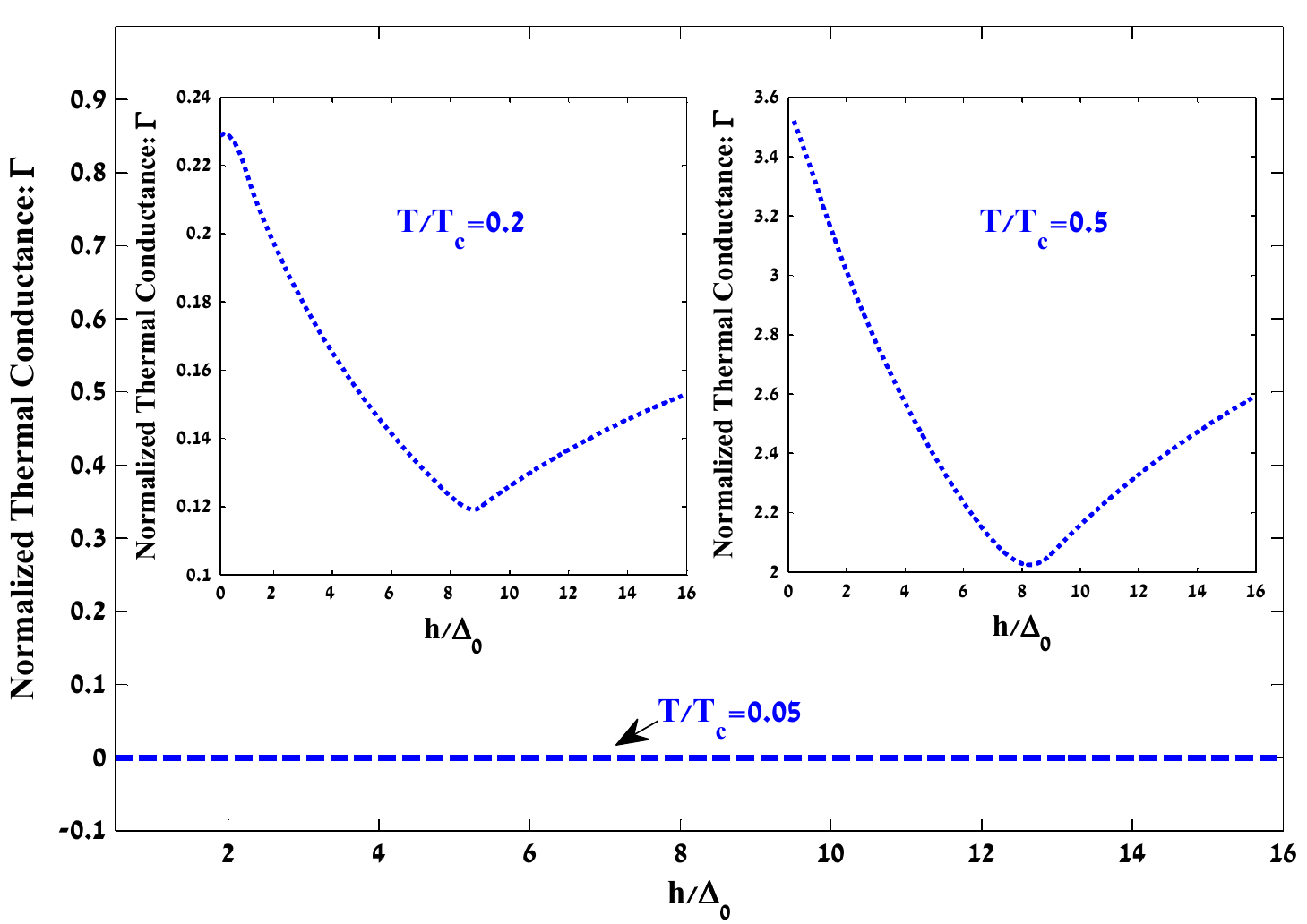}
\caption{\label{fig:KapavshSF} (Color online) The normalized
thermal conductance of the single FG/SG junction vs exchange field
of the ferromagnetic layer for three values of temperatures
.}
\end{figure}
where $\Gamma_0=E_F/2\pi^{2}\hbar^{2}v_{F}k_{B}\Delta_0$ is a
constant. Here all parameters are normalized, i.e. energies, with
respect to pair potential at zero temperature
($\Delta_0\equiv\Delta(T=0)$) , and temperatures, with respect to
the critical temperature of superconducting order parameter ($Tc$).
 Throughout the paper we
set $\Delta_0=\hbar=k_B=1$ in our computations.
\section{\label{sec:results}Results and Discussion}
Now we proceed to present the main results of the paper. In order to
obtain more realistic experimental results, we consider highly
doping junctions and set a large mismatch potential $U_0$. The Fermi
energy of the graphene is externally controllable and can be
tuned\cite{beenakker1,beenakker3}. Energy of quasiparticles remains
within our regime of validity, for values near 0 to 1eV i.e. the
quasiparticles follow the Dirac equation. Throughout the paper we
fix the Fermi energy $E_F=10\Delta_0$ which typically places Fermi
energy within 10-15 meV. We change the width of the junction from
$L\simeq\xi_S$ up to $L\simeq9\xi_S$ and investigate how the thermal
conductance is varied by the variation in $L/\xi_S$. Also we
investigate other possibilities of variations in every parameter
involved the problem and how they influence the thermal conductance.
When we need fixed temperature, use $T/Tc= 0.2$, when we need fixed
exchange field use $h/\Delta_0=8$ and when we need a fixed width,
typically use $L/\xi_S=4$. As will be discussed in the following
sections, we find that unlike the exponential increase of the
normalized thermal conductance with respect to  the temperature in
the FG/SG junctions reflecting
 the $s$-wave superconducting correlation, the normalized thermal conductance for FG/SG/FG junctions shows a
linear increasing with respect to the temperature for small widths
($L\simeq\xi_S$) of the superconducting region \textit{i.e.}
$\Gamma\propto T$. For FG/SG/FG junctions, the thermal conductance
vs. width of the superconducting region $L/\xi_S$ shows an intensive
oscillatory behavior, also it has a minimum vs. strength of the
exchange field in both semi-infinite ferromagnetic graphene
sheets. \\
\begin{figure}
\includegraphics[width=8.50cm]{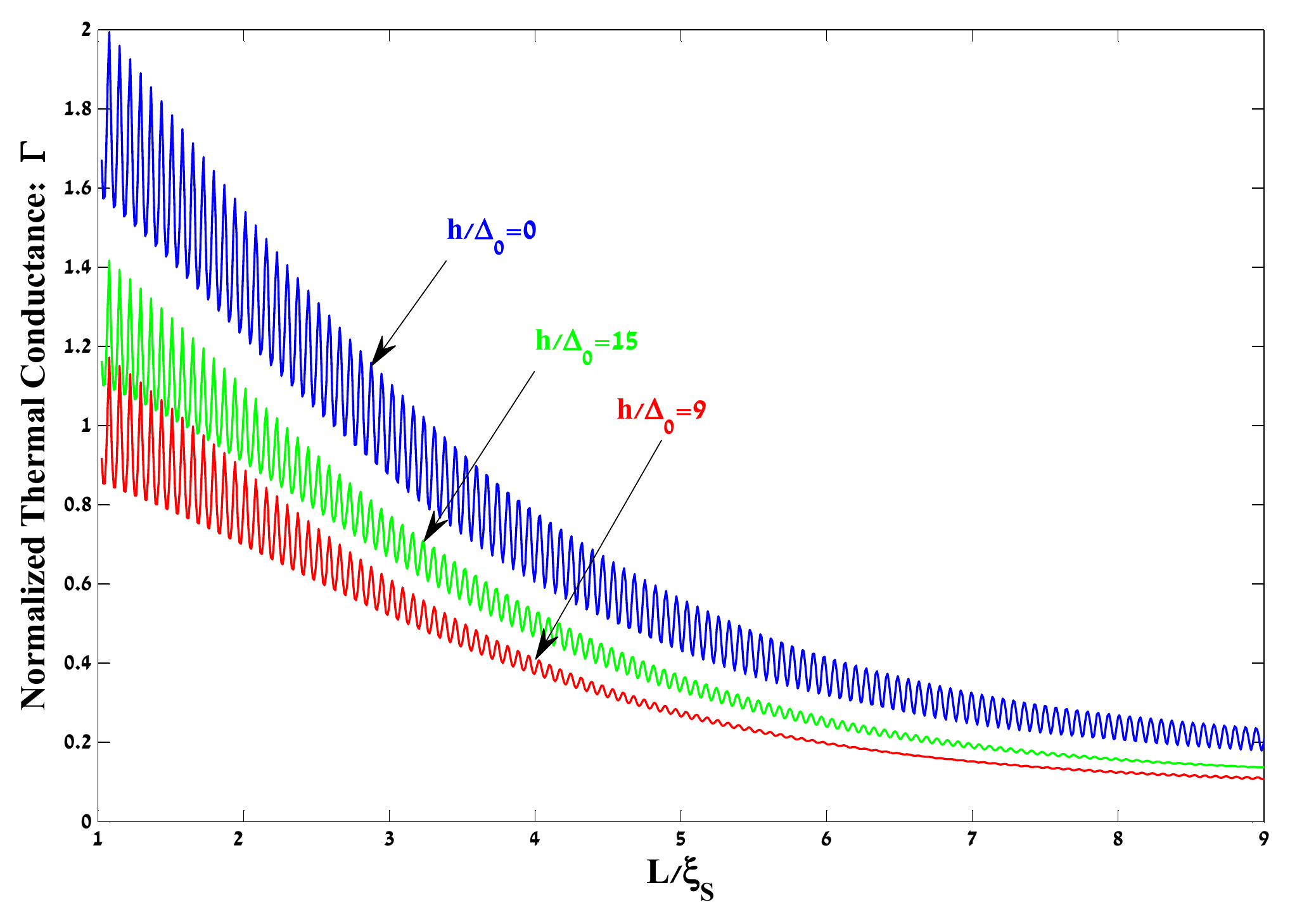}
\caption{\label{fig:kapavsL} (Color online) The normalized thermal
conductance $\Gamma$ vs the normalized width $L/\xi_S$ of
graphene-based superconducting layer for different values of $h$
in FG/SG/FG junction.
  The temperature is fixed at $T/T_c=0.2$.}
\end{figure}
\subsection{Heat transport through the single
Ferromagnetic/Superconductor junction} At first, we investigate how
the normalized thermal conductance of the single junctions (FG/SG)
behavior by changing both the temperature and exchange field $h$ of
the semi-infinite ferromagnetic sheet.
 We fix the Fermi levels
of superconductor and ferromagnetic sheet at $E_F=10\Delta_0$ and
use a large mismatch potential $U_0$, thus with such parameters we
remain in the heavily doped regime.
 As seen in Fig. 2, the thermal conductance increases exponentially by increasing the temperature T, that verify
the presence of the induced s-wave correlation  in the graphene sheet.
 As seen in Fig.\ref{fig:KapavsTSF}, the curve of the thermal conductance for $h/\Delta_0=15$
is placed between that of the thermal conductance for $h/\Delta_0=0
$ and 9. This findings show that the thermal conductance has a
minimum vs. $h/\Delta_0$ as seen in Fig. \ref{fig:KapavshSF}. In
Fig. \ref{fig:KapavshSF} the normalized thermal conductance reach to
a minimum near $h\simeq E_F$. Therefore by increasing the exchange
field of the ferromagnetic layer, propagating modes of the Dirac
Fermions decay , and reach to a minimum number near Fermi level.
 By increasing the temperature, the order parameter
of the superconducting correlations decays, thus increasing
of the temperature helps the exchange field to make  propagating
modes reach to their minimum value at smaller values of
exchange field $h$ as seen in Fig. \ref{fig:KapavshSF}.
 Now we proceed to present our findings for
double FG/SG/FG junctions.

\begin{figure}
\includegraphics[width=9.0cm]{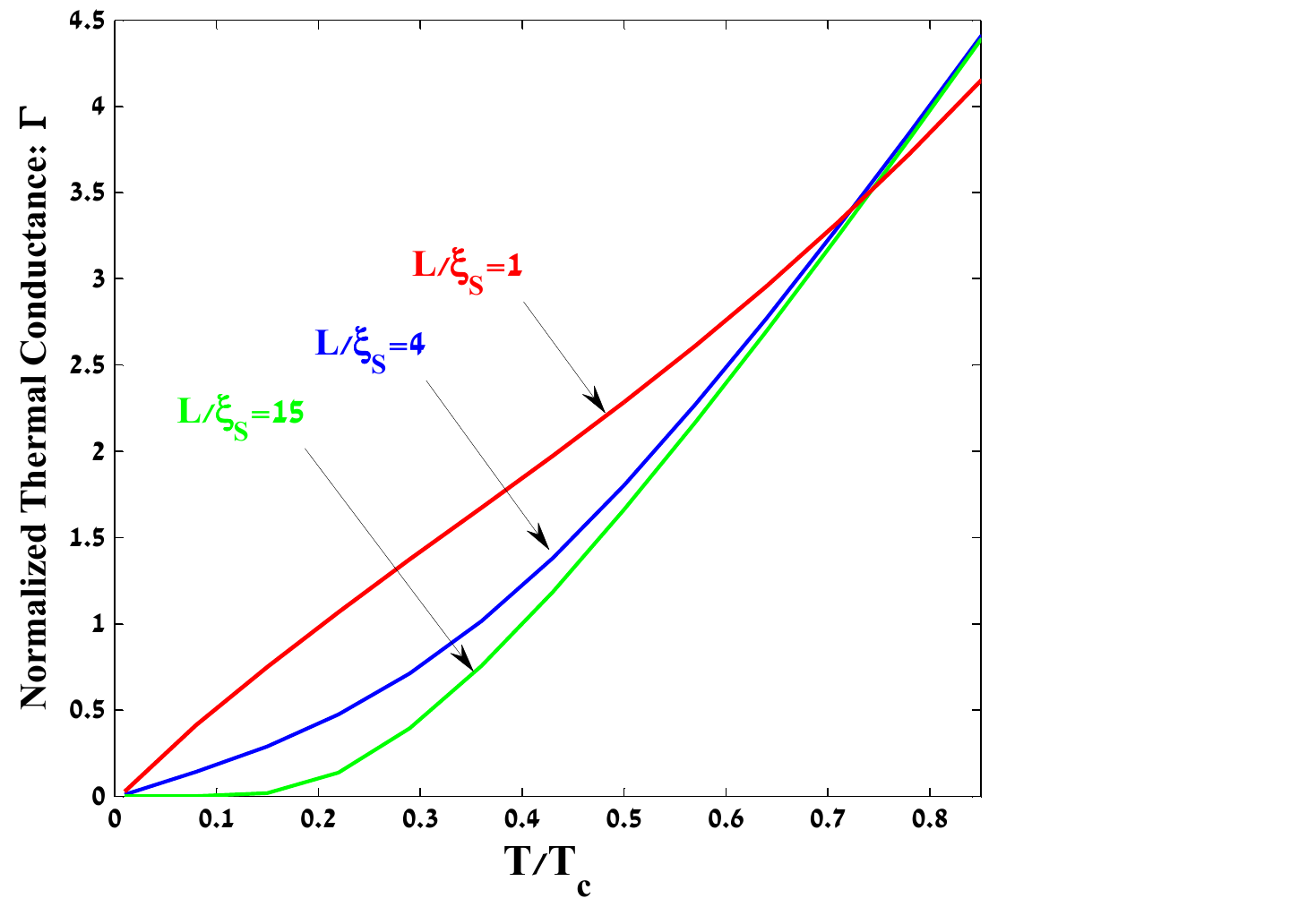}
\caption{\label{fig:kapavsT} (Color online) The normalized thermal
conductance $\Gamma$ vs temperature for three values of normalized
widthes of the superconducting layer $L/\xi=1, 4, 15$ in
FG/SG/FG junction. 
The exchange field is fixed at $h/\Delta_0$=8.}
\end{figure}

\subsection{Heat transport through the double
Ferromagnetic/Superconductor/Ferromagnetic junction } 
\begin{figure}
\includegraphics[width=10.5cm,height=7.5cm]{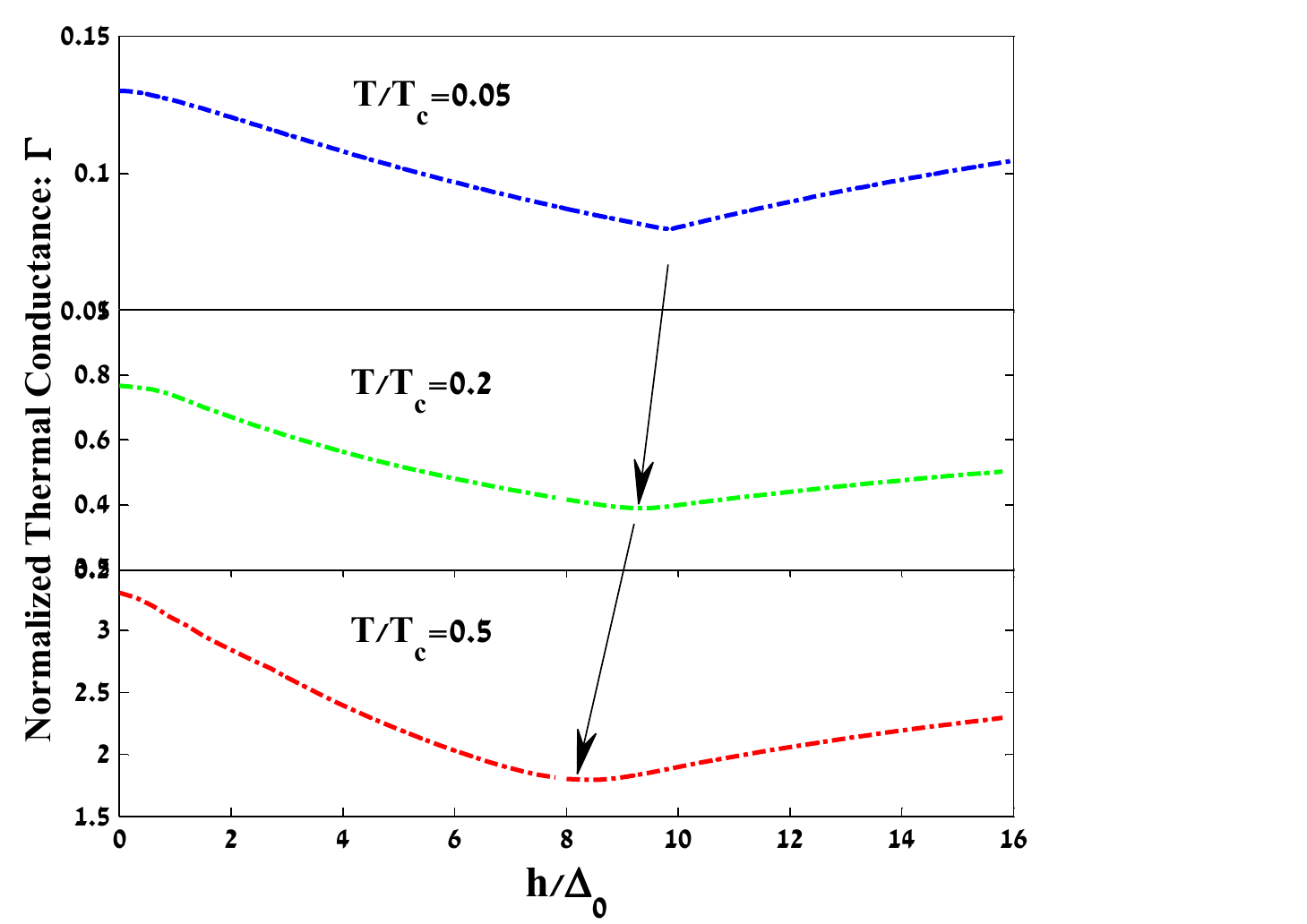}
\caption{\label{fig:kapavsh} (Color online) The normalized thermal
conductance $\Gamma$ vs strength of normalized exchange field
$h/\Delta_0$ of the ferromagnetic sides for three values of
temperatures in
FG/SG/FG junction. 
The width of the superconducting layer is fixed at $L/\xi_S$=4.}
\end{figure}

Now we turn
our attention to the second graphene-based structure, namely double
FG/SG/FG junction. The suggested set-up of the junction
schematically is shown in Fig. \ref{fig:model}. Such junctions can
be made by depositing a superconducting electrode on top of the
graphene sheet. As before we fix the Fermi energy of the
superconductor and ferromagnetic at $E_F=10\Delta_0$. As shown in
Fig.\ref{fig:kapavsL}, the normalized thermal conductance shows
intensive oscillatory behavior versus width of superconductor
region. This manner is a consequence of the coherent interference of
the Dirac fermions in facing with two interfaces. Also, by
increasing the width of superconducting region, the normalized
thermal conductance loses the amplitude of its oscillation and shows
exponentially decrement towards single junction. When the Dirac
fermions lose their coherency , they show the mentioned manner and
this illustrates the fact that when the width of the superconducting
region becomes larger, the probability of tunneling of the Dirac
fermions across the superconducting region is reduced. The thermal
conductance versus temperature is shown in \ref{fig:kapavsT}. The
thermal conductance for small widths of the superconducting region
shows a linear behavior similar to the behavior of the thermal
conductance of metals, discussed in the introduction.
 This manner reflects the tunneling process of the
Dirac fermions through the superconducting region. Thus one can
expect that when the width of this region goes to larger values,
this phenomena (tunneling) is reduced and the thermal conductance
behaves like the single junction which is clearly shown in
Fig.\ref{fig:kapavsT} for $L/\xi_S=15$. The thermal conductance as a
function of the exchange field of the ferromagnetic region is
plotted in Fig. \ref{fig:kapavsh}. The behaviors of the thermal
conductance vs. h for double junction (FG/SG/FG) and the single
junction (FG/SG) are the same. The thermal conductance reach to a
minimum near $h\simeq E_F$,and the minimum value is moved to smaller
ones by the temperature increment.
\section{Summary}
In the present paper we have considered two types of gapless
graphene-based junctions, FG/SG and FG/SG/FG with $s$-wave
superconducting electrodes, and we have investigated the heat
transport properties of the systems. In particular, we have
investigated how the variations of the variable quantities in the
system can influence the thermal conductance of the junctions.
Utilizing the Dirac-Boguliobov de Genne equation for quasiparticles
inside the graphene and appropriate boundary conditions for the
obtained wave functions, we have derived the Andreev and normal
reflection coefficients, and used them for calculating the
normalized thermal conductance numerically. We found that for single
junction (F/S) the normalized thermal conductance indicates the
previous exponential form vs. temperature. Increasing exchange field
$h$ lowers the propagating modes of the Dirac Fermions down to a
minimum value. Due to the tunneling phenomena of the Dirac fermions
through the superconducting layer, the heat transport properties of
the double F/S/F junctions are different from single F/S junction.
The thermal conductance vs temperature exhibits linear behavior in
the tunneling limit (small widths of the superconducting region).
The thermal conductance vs. width of the superconducting region
exhibits very intensive oscillatory behavior and also the quantity
indicates previous behavior of the single junction (F/S) vs. $h$
\textit{i.e.} the enhancement of the exchange field from $0$ up to a
value near $E_F$, lowers the propagating modes of the Dirac
fermions.
\section*{Acknowledgments}
The authors appreciate very useful and fruitful discussions with
Jacob Linder. The authors would like to thank the Office of Graduate
Studies of Isfahan University.
\bibliographystyle{aip}

\end{document}